\documentstyle[multicol,epsfig,prl,aps]{revtex}

\ifpreprintsty
 \def\multb{ }
 \def\multe{ }
 \else
 \def\multb{ \begin{multicols}{2}}
 \def\multe{ \end{multicols}}
 \fi

\begin{document}
\title{Competitions in layered ruthenates: ferro- vs. antiferromagnetism and
triplet vs. singlet pairing}
\author{I.I. Mazin and D.J. Singh}
\address{Code 6691, Naval Research Laboratory, Washington, DC 20375}
\date{\today}
\maketitle

\begin{abstract}
Ru based perovskites demonstrate an amazing richness in their magnetic
properties, including 3D and quasi-2D ferromagnetism, antiferromagnetism,
and unconventional superconductivity. Tendency to ferromagnetism, stemming
from the unusually large involvement of O in magnetism in ruthenates,
leads to ferromagnetic spin fluctuations in Sr$_2$RuO$_4$ and eventually
to $p$-wave superconductivity. A related compound Ca$_2$RuO$_4$ was
measured to be antiferromagnetic, suggesting a possibility of 
antiferromagnetic fluctuations in Sr$_2$RuO$_4$ as well. Here we report
first principles calculations that demonstrate that in both compounds 
the ferro- and antiferromagnetic fluctuations coexist, leading to an
actual instability in Ca$_2$RuO$_4$ and to a close competition between
$p$-wave and $d$-wave superconducting symmetries in Sr$_2$RuO$_4$. The 
antiferromagnetism in this system appears to be mostly related with
the nesting, which is the strongest at ${\bf Q}\approx (2\pi/3a,2\pi/3a,0)$.
Surprisingly,
for the Fermiology of Sr$_2$RuO$_4$ the $p$-wave state wins over the
 $d$-wave one everywhere except in close vicinity of the antiferromagnetic
instability. The most stable state within the $d$-wave channel has 
vanishing order parameter at one out of three Fermi surfaces in 
Sr$_2$RuO$_4$, while in the $p$ channel its amplitude is comparable 
at all three of them. 
\end{abstract}

\multb
In the last few years an understanding emerged, thanks to the progress in
the high-$T_{c}$ problem, that $s$-wave and $d$-wave pairing are not
entirely antagonistic. It is possible to obtain either symmetry in the
framework of one and the same model, depending on the actual values of
parameters. It is, however, commonly believed that the
triplet ($p$) and singlet ($s,d$) pairings are so different in their nature
that more than just changing numerical parameters of a model is needed to
switch between these symmetries. Here we show that a
realistic model of the superconducting layered ruthenate Sr$_{2}$RuO$_{4}$
is in fact unstable with respect to both $p$- and $d$-wave pairing and it is
a close competition between the two that determines the actual ground state.

It was suggested a few years ago that Sr$_{2}$RuO$_{4}$ may be a $p$-wave
superconductor\cite{SR}.
The main
consideration was that the sister 3D compound, SrRuO$_{3},$ is a strong
ferromagnet, so one could expect substantial ferromagnetic spin fluctuations
in Sr$_{2}$RuO$_{4}.$ It was also known that superconductivity in the
canonical triplet superconductor, $^{3}$He, is due to ferromagnetic spin
fluctuations, so it was natural to conjecture that the pairing in Sr$_{2}$RuO%
$_{4}$ was also triplet.
At that time there was hardly any experimental
evidence and no microscopic calculations to support this idea.
Since then convincing experimental evidence has
been collected (for review, see Ref. \cite{S-her}) that the
superconductivity in Sr$_{2}$RuO$_{4}$ is indeed
unconventional (not $s$-wave) and
most likely triplet.
 Microscopic calculations revealed the mechanism for
ferromagnetism in SrRuO$_{3}$\cite{singh96,mazin97} and
demonstrated that a tendency
to ferromagnetism is still present in Sr$_{2}$RuO$_{4},$ although it is
weaker and does not result in an actual magnetic instability\cite{mazin97b}.

The recent discovery of {\it anti}ferromagnetism in Ca$_{2}$RuO$_{4}$ forces us
to reconsider this simple picture. This compound differs from Sr$_{2}$RuO$%
_{4}$ only in that the RuO$_{6}$ octahedra are tilted and rotated\cite
{braden98}, as it is common in perovskites (cf. La$_{2}$CuO$_{4}).$ This
causes some modification of the hopping amplitudes compared to Sr$_{2}$RuO$%
_{4},$ but this modification has a tremendous effect on magnetic properties:
the material becomes antiferromagnetic with a substantial ($>1$ $\mu _{B}$/Ru)
magnetic moment.
This suggests that there should 
be some latent tendency to antiferromagnetism in Sr$_{2}$RuO$_{4}$ itself,
and casts doubt on the basic assumptions of Refs.\cite{SR,mazin97b} and others
that the spin fluctuations in this material are predominantly
of the ferromagnetic type.
To answer this question one cannot rely upon analogies with other materials,
but needs quantitative (at least semiquantitative) calculations.

Fortunately, and unlike cuprates and most 3$d$-oxides, the conventional local
density approximation (LDA) provides a very good description of magnetic
properties of the ruthenate-based perovskites. Previously we have performed
calculations\cite{singh96,mazin97}
for various ruthenates for which crystal structure
and magnetic properties are known experimentally, and we found excellent
agreement with the experiment: SrRuO$_{3}$ comes out ferromagnetic with the
total magnetization  1.59 $\mu _{B}/$f.u. (experiment: 1.6), CaRuO$_{3}$ is a
paramagnet on a verge of ferromagnetism, and the double perovskite Sr$_{2}$%
RuYO$_{6}$ is antiferromagnetic with 3 $\mu _{B}/$f.u., again in
accord with the experiment. Finally, in Sr$%
_{2} $RuO$_{4}$ the paramagnetic state comes out more stable in the LDA
calculations than either ferro- or antiferromagnetic one.

The reason for ferromagnetism in SrRuO$_{3}$ (and near-ferromagnetism in
CaRuO$_{3})$ is now well understood\cite{mazin97}: There is substantial
oxygen density of states at the Fermi level in these ruthenates (due
 to strong $p-d$ hybridization), and the
difference between the FM and the AFM state is that in the latter case the
oxygen is not spin-polarized. The oxygen ion has considerable Stoner energy,
which is lost in the AFM case. Furthermore, this additional energy is
entirely lost for the ${\bf q}=\{\pi ,\pi ,\pi \}$ AFM ordering, two thirds
of that is lost for ${\bf q}=\{\pi ,\pi ,0\},$ and one third for ${\bf q}%
=\{\pi ,0,0\},$ compared with the FM ordering 
(${\bf q}=\{0,0,0\}).$ This allows one to construct a $q$-dependent Stoner
interaction, $I(q)\approx 0.46$~eV/$(1+0.08q^{2}),$ where $q$ is measured in
units of $\pi /a.$ This interactions strongly favors a FM instability, and
whether or not the actual instability occurs depends on the density of
states at the Fermi level, according to the Stoner criterion, $I(0)N(0)>1.$
It appears that in SrRuO$_{3}$ this condition is satisfied, $IN=1.23,$ and
the material is a FM. In CaRuO$_{3}$ the smaller ionic radius of Ca leads to
a smaller Ru-Ru distance and thus to larger distortion. A peak in the
density of states that exists in SrRuO$_{3}$ is washed out and the material
is on the border line, $IN\approx 1.$

The same mechanism is operative in Sr$_{2}$RuO$_{4}:$ for an individual RuO$%
_{2}$ plane we obtain a (2D) Stoner factor $I(q)\approx 0.43$~eV/$%
(1+0.08q^{2}),$ favoring ferromagnetic spin fluctuations in the plane\cite
{mazin97b}. However, the 2D character of the band structure of Sr$_{2}$RuO$%
_{4} $ introduces additional complications. As discussed in Refs. \cite
{DJS,mazin97b}, of the three Fermi surface sheets one ($\gamma $) is
quasi-isotropic 2D, and two ($\alpha $ and $\beta $) are quasi-1D. The
latter can be visualized (cf. Fig.1 in Ref.\cite{mazin97b}) as a system of
parallel planes separated by $Q=2\pi /3a,$ running both in the $x$ and $y$
directions. This is true in the nearest neighbor $dd\pi $ tight-binding
model, while in reality due to the next hoppings the planes are warped and
reconnected at the crossing lines, to form two pseudo-square prisms,
obtained in LDA calculations and 
observed experimentally. Naturally, such a Fermi surface should give rise
to sizable nesting effects at the wave vectors ${\bf k}=(Q,k_{y}),$ ${\bf k}%
=(k_{x},Q),$ and especially at ${\bf k=Q}=(Q,Q).$ This would lead to AFM
spin fluctuations at these vectors, in addition to the FM fluctuations
discussed above. To check, we have integrated the LDA band structure of
Sr$_{2}$RuO$_{4}$ to get the bare RPA susceptibility, 
\begin{equation}
\chi _{0}({\bf q)}=\sum_{{\bf k}ij}\frac{M_{{\bf k}i,{\bf k+q}%
,j}[f(\varepsilon _{{\bf k},i})-f(\varepsilon _{{\bf k+q},j})]}{\varepsilon
_{{\bf k+q},j}-\varepsilon _{{\bf k},i}},  \label{chi0}
\end{equation}
where $f$ is the Fermi distribution function, $i$ and $j$ label the three
bands. All $|{\bf k}i>$ states were classified according to the maximal $%
t_{2g}$ character, $xy,$ $yz,$ and $zx,$ and the matrix element $M$ is taken
to be 1 between two states which have the same maximal character and 0
otherwise. This is, of course, a rather crude approximation, but it should
reveal the qualitative behavior of $\chi _{0}.$ The results are shown in
Fig.1. Roughly speaking, 
\begin{equation}
\chi _{0}({\bf q)}=N(0)+\chi _{n}({\bf q}),  \label{model}
\end{equation}
where $\chi _{n}$ is the nesting-dependent contribution. The total
susceptibility can then be expressed as 
\begin{equation}
\chi ({\bf q)}=\frac{\chi _{0}({\bf q)}}{1-I(q)\chi _{0}({\bf q})}=\frac{%
\chi _{0}({\bf q})}{1-I(q)N(0)-I(q)\chi _{n}({\bf q})}.  \label{chi}
\end{equation}
This form implies two different kinds of spin fluctuations: FM ones, {\bf q}$%
=0,$ and AFM ones, at ${\bf q}={\bf Q}$. If $I(Q)N(0)+I(Q)\chi _{n}({\bf Q}%
)>I(0)N(0),$ the AFM fluctuations are stronger. This seems to be the case in
Sr$_{2}$RuO$_{4}:$ our calculations yield $I(0)N(0)=0.82,$ in good agreement
with the experimentally observed susceptibility enhancement, and $%
I(Q)N(0)+I(Q)\chi _{n}({\bf Q})=1.02$ (which actually corresponds to an
instability with respect to tripling of the unit cell both in $x$ and in $y$%
). Since no instability is observed in the experiment, nor in the direct
calculations, we conclude that the approximate treatment of the matrix
elements in Eq.\ref{chi0} leads to an overestimation of $\chi _{n}$ by at
the very
least 2\%, but the conclusion that 
AFM fluctuations are stronger or at least comparable
with the FM ones likely holds.
\begin{figure}[tbp]
\centerline{\epsfig{file=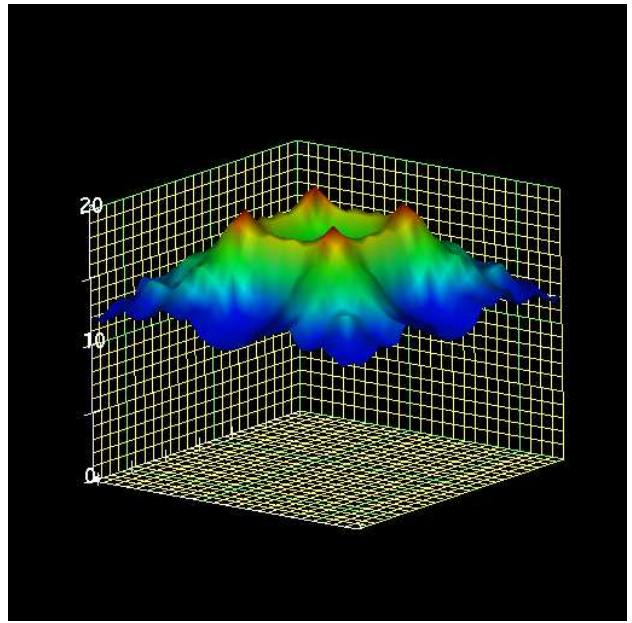,width=0.95\linewidth}}
\vspace{0.in} \setlength{\columnwidth}{3.2in} \nopagebreak
\caption{Calculated bare susceptibility for Sr$_2$RuO$_4$}
\label{chifig}
\end{figure}

The direct way to test this experimentally is via neutron
scattering\cite{imainote}. There is, however, an indirect argument in favor
of strong AFM spin fluctuations. Increasing the effective dimensionality by
adding additional RuO$_{2}$ layers, one can increase $N(0)$ and eventually
get a FM instability. Experimentally this happens when the number of layers
is 3 or maybe even 2\cite{note}. Another possible (but not guaranteed)
effect of adding layers is increased $z$-dispersion and thus deteriorated
nesting. On the other hand, reducing the next-nearest-neighbor hopping should
improve nesting and make an AFM transition more likely. One expects such a
reduction from rotating the RuO$_{6}$ octahedra\cite{rot}, as 
for instance in Ca$_{2}$RuO$_{4}.$ Indeed, experimentally Ca$_{2}$RuO$_{4}$
 is an
AFM with a magnetization of 1.2--1.3 $\mu _{B}$/Ru
 and $T_{N}\approx 150$ K. Moreover,
this AFM state is remarkably different from typical Mott-Hubbard insulators,
driven by strong Coulomb correlations. First, although the conductivity
grows with temperature, the functional dependence is consistent with a
variable-range hopping and not with activation. Second, there is substantial
density of states at the Fermi level,
as evidenced by specific heat
measurements. These two facts indicate that Ca$_{2}$RuO$_{4}$ is not a
simple
insulator, but a metal with disorder localized carriers (which is in turned
helped by strong coupling between the spin and charge degrees of freedom
\cite{mazin97}). 
We performed LDA calculations for Ca$_{2}$RuO$_{4}$ 
similar to those reported in Refs.\cite{singh96,mazin97}and found a
magnetic moment of  $\approx 1.5  \ 
\mu _{B}$ (of which  $\approx 1 \ \mu _{B}$ is inside the the Ru MT
sphere and the rest mostly residing on the apical oxygens) for Ca$_{2}$RuO$%
_{4}$ (in agreement with the experimental $%
M=1.3\ \mu _{B}$). In fact, we also find a FM instability, and that
the FM and the AFM states Ca$_{2}$RuO$
_{4}$ degenerate within the LDA to
within  a few meV/atom, indicating a close competition between these two
magnetic states\cite{note3}. For genuine Mott insulators the LDA either fails
to reproduce the magnetic instability, or underestimates the magnetization.
The calculated density of states is sizable, $N(E_F)=1.6$ st./eV spin f.u.),
but the corresponding effective mass is large
 (the in-plane average\cite{mass} $n/m$= $7.5\times 10^{20}$ cm$^{-3}/m_0$),
because Ru is strongly spin polarized and thus Ru-Ru nearest neighbor
hopping is suppressed with AFM order. This
also facilitates localization.

\begin{figure}[tbp]
\centerline{\epsfig{file=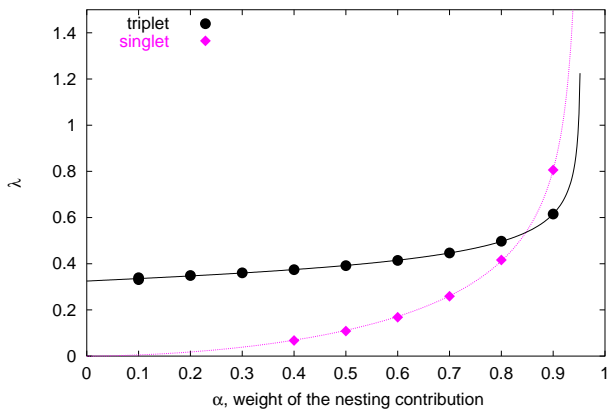,width=0.95\linewidth}}
\vspace{0.in} \setlength{\columnwidth}{3.2in} \nopagebreak
\caption{The maximum eigenvalue for the coupling matrix in the
singlet and in the triplet channels, as a function of the relative
strength of the AFM component in the bare electronic susceptibility.}
\label{lamfig}
\end{figure}

We now enumerate the relevant magnetic interactions in the (Sr,Ca) $%
_{2}$RuO$_{4}$ system. The first two have been discussed above, and they are the
Stoner ferromagnetism (${\bf Q=0}),$ nesting-derived antiferromagnetism ($
{\bf Q}\approx \{2\pi /3,2\pi /3\}).$ As usually, there is also
 superexchange $({\bf Q}
=\{\pi ,\pi \}).$ One should be reminded at this point, that superexchange 
{\it per se }is not a strong correlation phenomenon. In a quasi-one-electron
approach like LDA, for sufficiently large exchange splittings, an AFM state
is lower in energy than the FM one by roughly $Zt^{2}/\Delta ,$ where $Z$ is
coordination, $t$ is the hopping, and $\Delta $ is the exchange splitting.
Since $t$ is usually overestimated in LDA, and $\Delta $ is usually smaller
than Hubbard $U,$ the effect of superexchange is usually overestimated, not
underestimated in LDA (see Ref. \cite{mazin97} for more).

While superexchange is a universal mechanism and should also be
operative in Ca$_{2}$RuO$_{4},$ it hardly plays a leading role. Otherwise,
the LDA calculations that gave the right Ru moment would have overstabilized
the AFM state, while in reality the opposite is true (they come out
degenerate instead of the AFM being the ground state). The most likely
cause for the instability is nesting. Furthermore, superexchange is of less
relevance for Sr$_{2}$RuO$_{4}$ because it manifests itself for finite
amplitude spin fluctuations, and not in the low-energy spectrum.

Assuming that the  LDA gives a reasonable description of the spectrum of spin
fluctuations in Sr$_{2}$RuO$_{4},$ we can discuss the consequences that the
AFM spin fluctuations may have on superconductivity. 
Recall that in $p$-wave superconductivity only the small $q$ spin
fluctuations are pairing, while in the $d$-wave case they are mostly
pair-breaking. For a spectrum with a complicated ${\bf q}$ dependence, as
the one described by Eq.\ref{chi}, the most favorable state is defined by
an interplay of the Fermi surface geometry and the structure of the
effective interaction. In the weak coupling near $T_{c}$ the gap equation
looks like 
\begin{equation}
\Delta _{{\bf k}i}=\sum_{{\bf k}^{\prime }j}V_{{\bf k}i,{\bf k}^{\prime
}j}\Delta _{{\bf k}^{\prime }j}, \label{gap}
\end{equation}
where for the singlet ($d$) pairing $\Delta $ is the order parameter, and $%
V_{{\bf k}i,{\bf k}^{\prime }j}$ is (negative) pairing interaction, which in
the same approximation as the one used in Ref.\cite{mazin97b} is given by 
\begin{equation}
V({\bf q=k-k}^{\prime }{\bf )}=-\frac{I^{2}(q)\chi _{0}({\bf q)}}{%
1-I^{2}(q)\chi _{0}^{2}({\bf q})}.
\end{equation}
For the triplet ($p$) pairing $\Delta $ is the amplitude of the order
parameter, usually denoted as $d,$ and $V_{{\bf k}i,{\bf k}^{\prime }j}$ now
includes the sign-changing angular factor: 
\begin{equation}
V({\bf q=k-k}^{\prime }{\bf )}=\frac{{\bf v}_{{\bf k}}\cdot {\bf v}_{{\bf k}%
^{\prime }}}{v_{{\bf k}}v_{{\bf k}^{\prime }}}\frac{I^{2}(q)\chi _{0}({\bf q)%
}}{1-I^{2}(q)\chi _{0}^{2}({\bf q})}
\end{equation}

The largest eigenvalue of the matrix $V$ in Eq.\ref{gap} defines the
critical temperature, and the corresponding eigenvector defines the
anisotropy of the gap near $T_{c}$. Correspondingly, the instability in the
singlet channel is defined by the same matrix, but with the opposite sign
(the interaction is repulsive in singlet channel), and without the angular
factor $\frac{{\bf v}_{{\bf k}}\cdot {\bf v}_{{\bf k}^{\prime }}}{v_{{\bf k}%
}v_{{\bf k}^{\prime }}}.$ We have solved these eigenvalue problems
numerically using a discrete mesh of {\bf k} points at the Fermi line. We
used model susceptibility (\ref{model}), adding an adjustable reduction $%
\alpha $ for the nesting part, $\chi _{0}({\bf q)}=N(0)+\alpha \chi _{n}(%
{\bf q}),$ and look for the solutions at different $\alpha <0.98$ ({\it i.e.}%
, below actual AFM instability). Furthermore, to simplify the calculations,
we used an analytical form for $\chi _{n}({\bf q})$ that recovers the main
qualitative characteristics of the numerical result of Fig.1, namely $\chi
_{n}({\bf q})=A[\cos (aq_{x})+\cos (aq_{y})]+B[\cos (2aq_{x})+\cos
(aq_{y})]+C[\cos (aq_{x})+\cos (aq_{x})].$

The results are shown in Fig. 2. Amazingly, only in close vicinity to the
actual AFM instability ($.85<\alpha <0.98)$  the superconducting transition
happens in the $d$-wave channel. For smaller $\alpha $ the $p$-wave state is
favored. An interesting question is, what is the angular and/or
interband anisotropy of the gap. We illustrate this anisotropy for the
critical value of $\alpha $ (Fig. 3). The  $p$-wave state is relatively
isotropic, indicating
 that at least in the Fermiology there is no
reason for substantial interband anisotropy (``orbital-dependent
superconductivity''), suggested in Ref.\cite{SR}. The $d$-wave,
to the contrary, is substantially anisotropic beyond the standard
anisotropy associated with the nodes along the $\{\pi,\pi\}$
direction, namely the so-called $\alpha$ pocket of the Fermi surface
 has nearly
vanishing order parameter.

\begin{figure}[tbp]
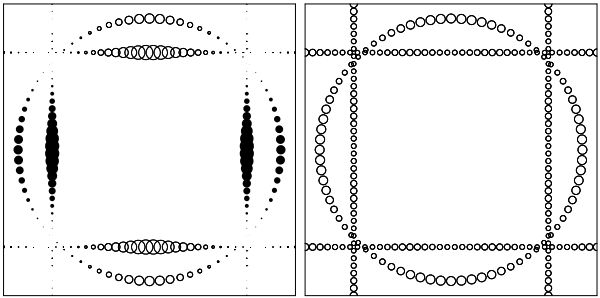

\centerline{\epsfig{file=s83.1.eps,width=0.45\linewidth} 
\epsfig{file=p83.1.eps,width=0.45\linewidth}}
\vspace{0.in} \setlength{\columnwidth}{3.2in} \nopagebreak
\caption{Relative magnitude of the order parameter in the
singlet (left) and the triplet (right) states. Parameter $\alpha$
is the same for both panels (0.83), corresponding to the 
coupling constant in both channels $\lambda\approx 0.5$.}
\label{sym}
\end{figure}
Our conclusions are as follows: (1) LDA calculations yield a self-consistent
solution with the AFM ordering for Ca$_2$RuO$_4$, degenerate with a FM
solution. The nonmagnetic state is considerable higher in energy, in agreement
with the experiment. The calculated magnetization value of $\approx 1$ $\mu_B$
is in agreement with the experiment as well. (2) This fact suggests that
besides the known tendency to ferromagnetism, there is an intrinsic
tendency to antiferromagnetism in Ru based layered perovskites. (3) Analysis of
the Fermi surface geometry of Sr$_2$RuO$_4$,
 as well as direct calculations, indicate
a strong nesting at ${\bf Q}=\{\pm 2\pi /3,\pm 2\pi /3\}$, as well
as a weaker nesting for the wave vectors connecting these four points
(Fig. 1). This result also suggests that the AFM ordering in Ca$_2$RuO$_4$
is mostly due to nesting, although superexchange may play some
role as well. It is, however, unlikely that this compound is a Mott-Hubbard
insulator. (4) The spin fluctuations in Sr$_2$RuO$_4$ have both a FM and 
an AFM component, of comparable
magnitude, thus making $d$-wave superconductivity
a strong competitor with the $p$-wave state. It is possible that the
system may be driven to the d-wave superconductivity by an external force,
for instance by pressure. (5) Spin fluctuation spectrum combined
with the actual fermiology of Sr$_2$RuO$_4$ produced a $p$-wave state
with little angular or interband anisotropy. (6) The sister compound
Ca$_2$RuO$_4$ is a low carrier density metal, where carrier are localized
due to disorder and 
electron-magnon scattering, and where antiferromagnetism is
primarily due to the nesting effects.

We acknowledge useful discussions with M. Sigrist, J.W. Lynn and E. Demler.
Computations were performed using facilities of the DoD HPCMO ASC Center.
Work at the Naval Research Laboratory is supported by the Office of the
Naval Research.


\begin{references}
\bibitem{rot} We do not discuss here the smearing of the density of 
states due to additional Bragg reflections in a distorted structure, 
an effect that controls the loss of magnetic ordering from Sr$Ru$O$_3$
to Ca$Ru$O$_3$.
\bibitem{mass} The effective mass has substantial (factor of 3)
 in-plane anisotropy.
\end{references}

\multe
\end{document}